**Reliability of Open Public Electric Vehicle Direct Current Fast Chargers**

David Rempel [1]

Carleen Cullen [1,2]

Mary Matteson Bryan [1]

Gustavo Vianna Cezar [3]

[1] Department of Bioengineering, University of California, Berkeley, CA, USA

[2] Cool the Earth, Kentfield, CA, USA

[3] SLAC National Accelerator Laboratory, GISMo Group, CA, USA

Word Count: 4476

Address correspondence to David Rempel, Department of Bioengineering, University of California, Berkeley, 1301 S. 46th Street, UC Berkeley RFS Building 163, Richmond, CA 94804, USA; e-mail: david.rempel@ucsf.edu.

## Abstract

In order to achieve a rapid transition to electric vehicle driving, a highly reliable and easy to use charging infrastructure is critical to building confidence as consumers shift from using familiar gas vehicles to unfamiliar electric vehicles (EV). This study evaluated the functionality of the charging system for 657 EVSE (electric vehicle service equipment) CCS connectors (combined charging system) on all 181 open, public DCFC (direct current fast chargers) charging stations in the Greater Bay Area. An EVSE was evaluated as functional if it charged an EV for 2 minutes or was charging an EV at the time the station was evaluated. Overall, 72.5% of the 657 EVSEs were functional. The cable was too short to reach the EV inlet for 4.9% of the EVSEs. Causes of 22.7% of EVSEs that were non-functioning were unresponsive or unavailable screens, payment system failures, charge initiation failures, network failures, or broken connectors. A random evaluation of 10% of the EVSEs, approximately 8 days after the first evaluation, demonstrated no overall change in functionality. This level of functionality appears to conflict with the 95 to 98% uptime reported by the EV service providers (EVSPs) who operate the EV charging stations. The findings suggest a need for shared, precise definitions of and calculations for reliability, uptime, downtime, and excluded time, as applied to open public DCFCs, with verification by third-party evaluation.

**Background**

Reliable, functional, open, public Direct Current Fast Charge (DCFC) electric vehicle (EV) charging stations are critical as countries rapidly transition to EVs. A recent survey of EV drivers in California (N=1290) reported mixed experience with existing EV chargers (CARB, 2022a). They reported experiencing broken plugs (9%), unexpected shut off during charging (6%), charging station not functioning (22%), payment problems (18%), and the need to contact customer service via cell phone (53%). This experience appears to contradict a simultaneous survey of the EV service providers (EVSPs) who reported 95 to 98 percent uptime of their public chargers.  An accurate assessment of the reliability, functionality, and uptime of the existing public EV chargers is needed to provide guidance for the successful buildout of the EV charging infrastructure.

Open EV charging stations are those open to all EVs (NREL, 2022).  Closed systems, such as Tesla Superchargers, will not accommodate all EVs.  Public charging stations are those that are open to the public 24 hours per day 7 days per week (AAI, 2022; NESCAUM, 2019). Examples of non-public charging stations are those in paid parking lots or those limited to customer and employee use. Open, public DCFC charging stations are designed to charge different models of EVs and, therefore, have multiple connector types, such as CCS (Combined Charging System; SAE, 2018), CHAdeMO, and Tesla connectors. Charging stations have one or more kiosks (also called posts), with each kiosk situated adjacent to one or two parking spaces. A kiosk may have one or more EVSEs (Electric Vehicle Supply Equipment) or ports (OCPI, 2020). An EVSE or port provides power to charge only one vehicle at a time even though it may have multiple cables with the same or different connector type (Figure 1). The EVSE provides information on charging and controls the delivery of electricity to the cable (DOE AFDC, 2022). Each kiosk typically includes a payment system that collects payment information from credit cards, debit cards, membership cards or smartphone applications; the transaction may be by tap, insert, swipe, or near field detection depending on the payment method. Another method of payment is *Plug and Charge* where the only action required is to plug in the EV and the EV is automatically identified and linked to a previously established payment method (ISO, 15118).

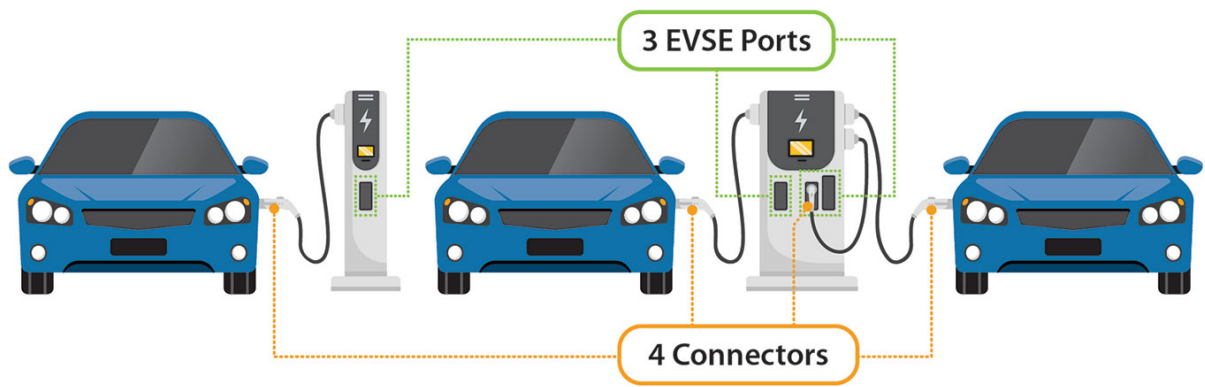

Figure 1.  A model of an EV DCFC charging station with 2 kiosks or posts, 3 EVSE charge ports, and 4 connectors.  Kiosks may have multiple connectors of the same or different types (e.g, CCS, CHAdeMO). [from DOE AFDC, 2022]

The National Renewable Energy Laboratory (NREL) Alternative Fuels Data Center (AFDC) maintains a national database/map of public EVSEs. The database includes charging station location and number of EVSEs (ports) and connection types at each station (NREL, 2022). The data is updated on a periodic basis by EV service providers (EVSPs); some states require updates at least monthly (CARB, 2022b). In addition, commercial smartphone, tablet, and desktop apps, such as PlugShare, provide EV users with information on the location of EV charging stations, the name of the EVSP, the number and types of connectors, the maximum power delivered, and other information.

There are different methods of measuring reliability of an electrical system, but essentially, it is the degree to which the performance of the system results in electricity being delivered to the customer in the amount desired (ORNL, 2004). The reliability of an EVSE, that is, the functional state, can be considered from the perspective of the EVSP or the EV driver. The EVSP may detect the state of an EVSE through its communication network, or as calls to a service number by EV drivers, as a measure of reliability. From the EV driver perspective, a reliable EVSE is one that charges the EV, for the expected duration, after using an appropriate payment method, at the expected rate (i.e., kW). The upper bound on charge rate is influenced by many factors including the EV's state of charge, the maximum rate allowed by the EV, and the charging station nominal rate. The Alliance for Automotive Innovation (2022) defines a reliability standard as one specifying a minimum uptime requirement. States have different minimum uptime requirements for EVSEs that are paid for with public funds. For the Northeast States (NESCAUM, 2019) "Each connector on each public DC fast charging station pedestal shall be operational at least 99 percent of the time based on a 24 hour 7-day week (i.e., no more than 1.7 hours of cumulative downtime in a 7-day period)." For California, "The equipment must be operational at least 97 percent of the standard operating hours of the charging facility for a period of 5 years" (CEC, 2021).

However, the use of *uptime* as the reliability metric is controversial since there is no standard definition nor is there a standard calculation methodology. Given the complexity of the EVSE ecosystem and technology stack, from hardware to software, ensuring a high uptime and assigning "uptime ownership" of each EVSE may be difficult and may require standardization across different jurisdictions.

The EVSE ecosystem is composed of different stakeholders. For example, when an EVSE is installed, it is connected to the local utility electrical infrastructure that delivers power to EVSE. The EVSE is installed by a certified installer, operated by the charge point operator (CPO) and located at a site where it may be owned and managed by a site host or the EVSP. The EVSE is connected to an internet service provider (ISP) network and a payment system. Finally, the EVSEs may be serviced by an EV servicing company.

Depending on the jurisdiction, the overall responsibility for keeping the EVSE functioning, can be either with the local electric utility, the installer, the site host, the CPO, or the servicing company. These stakeholders may be independent or may be integrated, i.e., installer can also be the CPO, etc. These stakeholders will likely have different levels of visibility over the status of the system. For example, the site host might have information about the electrical infrastructure and outages and physical damage to kiosks but not information about the functional status of each kiosk, whereas the CPO may have continuous EVSE status information. This partial visibility of the EVSE operation poses a challenge in maintaining a high uptime from the EV driver perspective. Moreover, since these stations are in public locations, events such as road blockage due to construction, theft, or vandalism can occur, which are beyond the immediate control of the CPO. Therefore, the complex nature of the ecosystem and the lack of a clear definition and metrics describing EVSE uptime may interfere with stakeholders' accountability.

For the purposes of this study, a *functional* EVSE is one that can charge for a minimum of 2 minutes, using an appropriate payment method, without the need to make a service call. An EVSE includes all the system components within a kiosk that are necessary for a successful charge, including the port, screen, network communication, payment system, power source, software, cable, and connector. If a kiosk has more than one cable with a CCS connector, the functionality of each connector is evaluated and reported as a separate EVSE.

The purpose of this study was to systematically evaluate whether open, public DCFC EV chargers with CCS connectors were functional in the 9 counties of the Greater Bay Area. California has the greatest density of public open DCFC chargers in the US (NREL, 2022) and within California the density is high in the Greater Bay Area.

**Methods**

All open, public DCFC EV charging stations with EVSEs with CCS connectors in the 9 counties of the Greater Bay Area were identified using the NREL NFDC database and the PlugShare.com website. Stations with CCS connectors with a charge rate >= 50kW were identified. The 9 counties were Alameda, Contra Costa, Marin, Napa, San Mateo, Santa Clara, San Francisco, Solano, and Sonoma. Non-open EV charging stations, e.g., Tesla, as well as non-public EV charging stations, e.g., stations in paid parking lots, private workplaces, or business sites with restricted access hours, were excluded.

The identified EV charging stations were visited by a driver with an EV with a CCS charge inlet. Each EVSE at the station was tested by plugging the CCS connector into the EV and attempting to initiate and sustain a charge for 2 minutes. If the charge was successful, the EVSE was classified as functional. The unique kiosk and CCS connector number or name were recorded. If the parking space was occupied by another EV and the EV was charging, the EVSE was classified as functional. If the parking space was occupied by a non-EV or by an EV and not charging, it was classified as not tested. If none of payment methods tested worked, or the EVSE was not functioning, or did not initiate or sustain a charge, the EVSE was classified as nonfunctional. If the cable was too short to reach the EV charge inlet, the EVSE was classified as a design failure.

The payment methods tested included 2 different functioning credit cards and the vendor mobile app or membership card. Payment methods were tested in the following order, credit card 1 insert, credit card 1 swipe, credit card 2 insert, credit card 2 swipe, then mobile app or membership card, until one of the payment methods was accepted. Each method, i.e, a swipe, was attempted twice before moving to the next payment method. The credit cards used for testing were Mastercard, Visa, and Amex. If any of the payment methods worked and led to a 2 minute charge, the EVSE was classified as functional. The EV drivers were instructed not to call the service number if the EVSE did not work; a functioning EVSE should not require a call to a service number.

Twenty volunteer EV drivers assisted in the testing of the EV charging stations. Only EVs with CCS charge inlets were used. The vehicles used for testing were the Chevy Bolt, Kia Niro, Hyundai Kona, Ford Mustang Mach E, and Porsche Taycan. The EV battery charge level was less than full at the time of testing. The volunteers were trained on the study methods and assigned EV charge stations to test. The survey was completed using a Qualtrics survey on a mobile device while the driver was at the charging station.

A random sample of 10% of the stations was tested at two points in time, approximately 1 week apart, to determine whether the functional state of the EVSEs changed over time.

**Results**

A total of 181 open public DCFC EV charging stations and 678 EVSEs with CCS connectors were identified in the 9 counties of the Greater Bay Area and visited between February 12, 2022 and March 7, 2022. Of these 678 EVSEs, in 21 instances, the adjacent parking space was occupied by a non-EV (7) or an EV that was not charging (14); therefore, these 21 EVSEs were excluded from the evaluation. The remaining 657 EVSEs that were evaluated are listed by EVSP in Table 1.

Table 1.  Evaluated open public DCFC EV charging stations and EVSEs by EV Service Provider

| EVSP | Stations | | EVSE[1] | |
|---|---|---|---|---|
| | N | % | N | % |
| ChargePoint | 23 | 12.7% | 44 | 6.7% |
| Delta | 2 | 1.1% | 3 | 0.5% |
| Electrify America | 54 | 29.8% | 379 | 57.7% |
| EV Connect | 2 | 1.1% | 3 | 0.5% |
| EVgo | 90 | 49.7% | 216 | 32.9% |
| Freewire | 2 | 1.1% | 2 | 0.3% |
| Greenlots | 1 | 0.6% | 2 | 0.3% |
| Powerflex | 3 | 1.7% | 4 | 0.6% |
| Volta | 4 | 2.2% | 4 | 0.6% |
| Total | 181 | 100.0% | 657 | 100.0% |

[1] An EVSE includes all the system components in a kiosk necessary to deliver a charge to a single connector.

*Reliability of EVSEs*

The functional states of the 657 EVSEs are summarized in Table 2. 72.5% of the EVSEs were functioning at the time of testing; 57.8% were tested and charged for 2 minutes and 15.4% were occupied by an EV that was charging. 22.7% of the EVSEs were not functioning. System electrical failures, e.g., screen blank or non-responsive, text on screen of "charger unavailable" or "connection error"; payment system failure; or charge initiation failure, were the most common causes of failure. A charge initiation failure occurred if the charge did not start after the payment was accepted or the charge started but was interrupted before 2 minutes of charging was completed. A payment system failure was recorded only after all payment methods were tested, each twice, and all failed. A broken connector, e.g., cracked or with bent pins, was recorded for 0.9% of EVSEs.

The cord was too short to reach the EV inlet for 4.9% (N=32) of EVSEs tested. This design failure was recorded at a ChargePoint station (1), EVgo stations (4), and Electrify America stations (27). The EVs tested were driven into the parking space either forward or backward during testing to position the EV inlet as close as possible to the charging kiosk. The EVs used, when it was recorded that the cord was too short, were all Chevy Bolts.

Table 2.  Functional states of 657 CCS DCFC EVSEs.

| | N | % |
|---|---|---|
| Functioning | | |
|     Charged for 2 minutes | 375 | 57.1% |
|     Occupied by EV and charging | 101 | 15.4% |
|     Total | 476 | 72.5% |
| Not Functioning | | |
|     Connector broken | 6 | 0.9% |
|     Blank or non-responsive screen | 23 | 3.5% |
|     Error message on screen[1] | 24 | 3.7% |
|     Connection error[2] | 7 | 1.1% |
|     Payment system failure[3] | 47 | 7.2% |
|     Charge initiation failure[4] | 42 | 6.4% |
|     Total | 149 | 22.7% |
| Station Design Failure | | |
|     Cable would not reach[5] | 32 | 4.9% |

[1] Charger error, unavailable, under maintenance, etc.
[2] Connection, network, communication error, etc.
[3] 12 of these were evaluated with 2 credit cards but not an app or membership card
[4] Short session failure
[5] At 3 EVSEs the space was too small to safely back into

*Reliability by EV Service Provider*

Three EVSPs, ChargePoint, Electrify America, and EVgo accounted for 97.3% (639 of 657) of the EVSEs evaluated. The functional states of the EVSEs for the 3 EVSPs are summarized in Table 3. It should be noted that most of the Electrify America kiosks each had 2 CCS connectors that were each tested and reported as independent EVSEs. However, the 2 CCS connectors could not be used simultaneously. If each of these kiosks were considered as a single EVSE, with functionality determined if either just one or both connectors provided a successful charge, the percent of functional EVSEs for Electrify America would have increased from 73.9 to 77.1%.

Table 3.  Functional State of EVSEs by the Top 3 EV Service Providers

|  | ChargePoint | | Electrify America | | EVgo | |
|---|---|---|---|---|---|---|
|  | N | % | N | % | N | % |
| Functioning |  |  |  |  |  |  |
|    Charged for 2-minutes | 21 | 47.7% | 228 | 60.2% | 120 | 55.6% |
|    Occupied by EV and charging | 6 | 13.6% | 52 | 13.7% | 37 | 17.1% |
|    Total | 27 | 61.4% | 280 | 73.9% | 157 | 72.7% |
| Not Functioning |  |  |  |  |  |  |
|    Connector broken | 0 | 0.0% | 2 | 0.5% | 3 | 1.4% |
|    Blank or non-responsive screen | 4 | 9.1% | 13 | 3.4% | 5 | 2.3% |
|    Error message on screen | 4 | 9.1% | 17 | 4.5% | 3 | 1.4% |
|    Connection error | 0 | 0.0% | 0 | 0.0% | 6 | 2.8% |
|    Payment system failure | 3 | 6.8% | 25 | 6.6% | 16 | 7.4% |
|    Charge initiation failure | 5 | 11.4% | 15 | 4.0% | 22 | 10.2% |
|    Total | 16 | 36.4% | 72 | 19.0% | 55 | 25.5% |
| Station Design Failure |  |  |  |  |  |  |
|    Cable would not reach | 1 | 2.3% | 27 | 7.1% | 4 | 1.9% |
| TOTAL | 44 | 100% | 379 | 100% | 216 | 100% |

*Payment Methods*

For the 375 EVSEs that charged for 2 minutes, the payment methods that worked are summarized in Table 4. The payment methods were tested in the order presented in Table 4. For example, 50.4% of the successful charges occurred after just the first credit card was inserted. However, 24.5% of the successful charges required an app or membership card for payment, i.e., attempts to pay with 2 credit cards were not successful.

Table 4.  Payment method that worked, in the order tested, for the 375 EVSEs that charged for 2 minutes.

|  | N | % |
|---|---|---|
| Credit card 1 insert | 189 | 50.4% |
| Credit card 1 swipe | 33 | 8.8% |
| Credit card 2 insert | 44 | 11.7% |
| Credit card 2 swipe | 8 | 2.1% |
| App or membership card | 92 | 24.5% |
| Free | 9 | 2.4% |
| Total | 375 | 100.0% |

*Testing EV Charging Stations at Two Points in Time*

Nineteen (19) randomly selected stations (88 EVSEs) were tested by 2 different EV drivers to determine if their functional state changed over time. The mean time between samplings was 8.0 days (SD=4.9). Eight of the EVSEs could not be compared between the time points because during one of the samplings the EVSE was occupied by a non-EV, an EV that was not charging, or the cord was too short. Of the remaining 80 EVSEs, 48 remained in a functional state, 14 remained in a non-functional state, and 18 (22.5%) changed state from functional to non-functional or a non-functional to functional (5 of these occurred with the same EV model). For the 14 EVSEs that remained in a non-functional state, the cause of failure was the same at both sampling times for 13 of them. The overall functional status changed little between the sampling times, i.e., 72.5% were functional at time 1 and 70.0% were functional at time 2.

**Discussion**

Of the 657 open public DCFC CCS EVSEs evaluated in this study, 72.5% were functional at the time of testing while 27.5% were either not functional or the cable was too short to reach the EV inlet. The most common cause of a nonfunctional EVSE was an electrical systems failure which included an unresponsive or unavailable screen, a payment system failure, a charge initiation failure, a connection failure, or a broken connector.

This is the first study we are aware of that systematically evaluated the functional state of open public EV chargers. The findings corroborate recent non-systematic surveys of EV owners. In a survey of 1290 EV owners, 34% reported that charging station operability issues were a barrier to using public charging stations (CARB, 2022a). In survey of 5500 EV owners, 25% of those who use public DCFCs reported a major difficulty with chargers being nonfunctional or broken (Plug In America, 2022). In the same survey, only 4% of Tesla owners reported a major difficulty with the Tesla closed DCFC system.

In the Greater Bay Area, 3 EVSPs, ChargePoint, Electrify America, and EVgo accounted for 97.3% of the 657 open public DCFC EVSEs evaluated. There were important functional and design differences between the stations installed by these EVSPs. ChargePoint had the highest percent of non-functional CCS EVSEs at 36.4% followed by EVgo (25.5%) and Electrify America (19.0%). The most critical design flaw was that 7.1% of the Electrify America cables were too short to reach the Chevy Bolt charger inlet, a problem that may be experienced by other EVs with the power inlet on the side of the vehicle. The cable length problem could be addressed with an industry standard on minimal cord length based on the kiosk location relative to the parking space.

The term reliability, when referencing an electrical system, typically refers to the percent of time, over a given time period, that the system is fully operational and able to deliver power at the intended level. This percent is also referred to as the *uptime*. For public EV charging stations, the definition from the Northeast States, is "the percent of time that a charging station must be functioning properly and available for use by EV drivers" and "Each connector on each public DC fast charging station pedestal shall be operational at least 99 percent of the time based on a 24 hour 7-day week (i.e., no more than 1.7 hours of cumulative downtime in a 7-day period)" (NESCAUM, 2019). New York, California, and the Federal Highways Administration require a minimum uptime of 97% (NYSERDA, 2021; CEC, 2021; FHWA, 2022).

The findings of this study suggest that the currently installed DCFC stations do not meet the 97 to 99% minimum uptime required by public funding agencies. The findings also appear to contradict the 95 to 98% national uptime levels reported by EVSPs (CARB, 2022a, p11). EVSPs do not report the details of how they define and calculate uptime. The EV charging infrastructure would greatly benefit from more data transparency and transparency on methodologies used by each EVSP in calculating uptime. For example, EVSPs could share data on the different subcomponent failure rates and whether the failure was localized, i.e., only affecting one EVSE due to a component failure, or systemic, i.e., affecting multiple EVSEs due to a communication or software problem. Such a reporting mechanism would benefit the entire industry by establishing an ongoing mechanism to identify the weak links in the ecosystem and developing a coordinated approach to addressing them.

While there are state reporting requirements for uptime; there are no precise state, national, or industry consensus definitions of nor calculation methods for uptime. A definition of uptime also requires a definition of the opposite, or *downtime*. Downtime is the total time that the EVSE is not operational. The clock on downtime should start when the EVSP has evidence that the system is unable to sustain a charge at the expected level. For example, recording downtime could start when there is (1) a system fault detected through the EVSP network where the fault results in the inability to charge, (2) a call to the service center by an EV driver to report non-functioning kiosk, (3) evidence of damage to physical components observed either in person or remotely, or (4) a nonfunctioning EVSE reported during a third-party evaluation of the station. If a failure is due to conditions outside of the control of the EVSP, e.g., upstream loss of power, cellular, or internet, it may be considered *excluded time*. If excluded time is used in calculating uptime, it should be subtracted from the reporting period time.

To improve the accuracy of reliability reporting, a third-party field audit of an EV charging station could be performed at the startup of the charging station and at periodic intervals thereafter. An audit of each EVSE should involve a standard methodology which could include an assessment of the allotted parking space, a measurement of the cable length, a test of payment methods and screen function, and a confirmation that power is delivered to the EV for a minimum period of time at the intended power level. A second type of third-party audit, following an Evaluation, Measurement and Verification (EM&V) process (DOE, 2022; CPUC, 2006), may also be useful to evaluate the EVSP system and data on uptime, downtime, and excluded time. Such audit findings should be made public.

To improve EV driver expectations and experience, accurate, real-time data on EVSE status should be made public. As mentioned before, the definition of reliability can be viewed from the perspective of the EV owner or the EVSE owner, and they are not necessarily the same. Acknowledging this difference, as the technology and regulatory framework matures and is better defined, is important to establish the correct expectations and prevent EV owners from giving up their EVs and returning to gas vehicles (Harding and Tal, 2021). Real-time data would allow EV owners to better understand the actual reliability of the EV infrastructure and adjust their expectations accordingly. Real-time data could be reported by EVSPs to the NREL Alternative Fuels Data Center (AFDC) and published on the National AFDC map and database. The data could also be made available for commercial applications that provide locations of EV charging stations and information on EVSE status to EV drivers.

Uptime may also be improved with standard maintenance and servicing agreements of EV charging stations. The Northeast State guidelines call for a 24-hour window for servicing an EVSE when the EVSE owner or operator is aware that an EVSE is not functioning (NESCAUM 2019). General maintenance may include the periodic checking of EVSE parts for damage; cleaning the EVSE kiosk, cables, and connectors; and removal of garbage and snow (NREL 2022).

Several limitations of the study should be noted. First, the test of functionality required a 2 minute successful charge of the EV. A charging process may be interrupted for no apparent reason at any time during charging, so the 2 minute duration may be too brief a test period to fully evaluate functionality. Second, the EV charging stations were evaluated at a single point in time, limiting conclusions about uptime. However, based on our reevaluation of 80 EVSEs, the functional state changed for 22.5% of the EVSEs, but the overall percent of functional EVSEs did not change. Third, the test method used different payments methods, 2 credit cards and an app or membership card. A well-functioning system should work with just one payment method. However, if the test methodology had required successful charging with just one credit card, the percent of functional EVSEs would have dropped from 72.5 to 49.2%. Fourth, the test methodology used did not include having the EV driver call a service number if they were unable to charge the EV. The need to call a service number for assistance might be considered by some a normally functioning system. Fifth, classifying "occupied by an EV and charging" as functional may overstate the overall percent functional since it is unknown whether the EV owner called the service number to initiate charging. Sixth, the test methodology did not determine whether the port was delivering power at the intended level; this should be included in future tests. Finally, the finding that the cable was too short to reach the EV inlet for 32 connectors is a major station design flaw. The identification of this problem was dependent on the EV model used for testing; testing with an EV that is not a Chevy Bolt may not identify this problem.

**Conclusions and Recommendations**

As more and more EVs are adopted nationally, the need for fully functional and reliable open public DCFCs will increase. Non-functional public chargers pose an important equity issue as residents in rented or multi-family dwellings usually charge at public charging stations. In addition, non-functional public chargers will have a significant impact on drivers on road trips. Furthermore, high rates of non-functional chargers may inhibit the adoption of EVs. The design of location and quantity of needed DCFC charging stations, for the build out of a national EV charge infrastructure, should not have to assume that a quarter of the EVSEs will be non-functional. The level of system failure observed indicates a poor quality of electrical design, components, or software plus the need for EVSPs to improve their identification of the EVSE functional status to trigger timely service. In addition, effective compliance measures are needed for EV charging stations that are part of a court settlement or paid for with public funds. Compliance measures require clear definitions of reliability, uptime, downtime, and excluded time. It may be useful to consider reliability metrics from other industries (e.g., data centers, cloud service providers, etc.), such as mean time to recovery or mean time between failures, etc. In addition, compliance measures may require third-party assessments of EVSEs, using a standard test methodology, at the time of initial operation and at regular intervals thereafter and an assessment of reliability data collected by the EVSPs.

**Acknowledgements**

We wish to thank the volunteer EV drivers who assisted in field data collection; these included Catherine Bohner, Suzanne Bryan, Lisa Chang, Ed Church, Jeff Cullen, Elena Engel, Ariane Erickson, CM Florkowski, Chris Gilbert, Howdy Goudey, Bill Hilton, Wiley Hodges, Linda Hutchins-Knowles, Douglas Mason, and Louie Roessler.  Partial funding for the study was provided by *Cool the Earth,* a 501(c)3 nonprofit organization.

The authors declare no financial conflict of interest.